\documentclass{aa}
\usepackage{psfig,graphicx}
\usepackage{txfonts}
\begin{document}

\title{Coma cluster object populations down to M$_R \sim -9.5$
\thanks{Based on observations obtained at the Canada-France-Hawaii Telescope 
(CFHT) which is operated by the National Research Council of Canada, the 
Institut National des Sciences de l'Univers of the Centre National de la 
Recherche Scientifique of France, and the University of Hawaii.}}

\author{C. Adami\inst{1} \and 
J.P. Picat\inst{2} \and 
F. Durret\inst{3,4} \and 
A. Mazure\inst{1} \and 
R. Pell\'o\inst{2} \and 
M. West\inst{5,6}
 }

\offprints{C. Adami}

\institute{
LAM, Traverse du Siphon, 13012 Marseille, France
\and
Observatoire Midi-Pyr\'en\'ees, 14 Av. Edouard Belin, 31400 Toulouse, France
\and
Institut d'Astrophysique de Paris, CNRS, Universit\'e Pierre et Marie Curie,
98bis Bd Arago, 75014 Paris, France 
\and
Observatoire de Paris, LERMA, 61 Av. de l'Observatoire, 75014 Paris, France
\and
Department of Physics and Astronomy,  University of Hawaii,
200 West Kawili Street, LS2, Hilo HI 96720-4091, USA
\and
Gemini Observatory, Casilla 603, La Serena, Chile
}

\date{Accepted . Received ; Draft printed: \today}

\abstract
{This study follows a recent analysis of the galaxy luminosity functions 
and colour-magnitude red sequences in the Coma cluster (Adami et al. 2007).}
{We analyze here the distribution of very faint galaxies and globular
clusters in an east-west strip of $\sim 42 \times 7$~arcmin$^2$
crossing the Coma cluster center (hereafter the CS strip) down to the
unprecedented faint absolute magnitude of M$_R \sim -9.5$.}
{This work is based on deep images obtained at the CFHT with the CFH12K
camera in the B, R, and I bands.}
{The analysis shows that the observed properties strongly depend on
the environment, and thus on the cluster history.  When the CS is
divided into four regions, the westernmost region appears poorly populated, 
while the regions
around the brightest galaxies NGC~4874 and NGC~4889 (NGC~4874 and NGC~4889
being masked) are dominated by faint blue galaxies. They show a faint 
luminosity function slope of -2, very significantly different from the 
field estimates.}
{Results are discussed in the framework of galaxy destruction (which
can explain part of the very faint galaxy population) and of
structures infalling on to Coma.}


\maketitle

\section{Introduction}

A few ultra deep (R of at least $\sim$25) luminosity functions (LFs)
have become available in the literature for the Coma cluster
(Bernstein et al. 1995: B95 or Milne et al. 2007: M07) but only in
small fields. Recent large scale surveys for Coma are limited to much
brighter magnitudes (e.g. Lobo et al. 1997, Trentham 1998, Terlevich
et al. 2001, Andreon $\&$ Cuillandre 2002, Beijersbergen et al. 2002,
Iglesias-P\'aramo et al. 2003).

Ultradeep LFs sample the Coma cluster population down to the limit of
what can be called galaxies and then include large populations of
globular clusters. This is also shown by M07, who considered a
catalog of objects detected along the Coma cluster line of sight down
to R$\sim$25.75. Among this population, the objects belonging to the
Coma cluster are M$_R \sim -9$ systems. This corresponds to galaxies
with a mass similar to that of globular clusters, and sometimes even
smaller. The formation of such systems (that would not be globular
clusters) remains quite puzzling and could e.g. be
related to formation processes of tidal dwarf galaxies (e.g. Bournaud
et al. 2003). This raises the question of galaxy formation in clusters
from material coming from other already existing galaxies and the
question of the globular cluster removal from their parent
galaxies. It also relates to the cluster ability to influence the
different giant galaxy types (e.g. Boselli et al. 2006) that will
contribute in different ways to the formation of these very faint
dwarf galaxies. A nice example of such very faint objects torn off
from galaxies infalling into massive clusters can be found in Cortese
et al. (2007).

To initiate the formation processes of such faint systems, material
needs to be expelled from existing galaxies by several internal
(e.g. gas expulsion via supernova winds) or external (tidal
disruptions, harrassment, ...)  processes. At least external processes
are driven by environmental effects that act at the cluster scale. In
order to test these scenarios, we therefore need surveys with such
large characteristic scales. However, up to now these ultra-deep
surveys were limited to very small areas (a few to a few tens of
arcmin$^2$ for M07 and B95 respectively), which is a limitation to
study possible environmental effects in the Coma cluster.

In order to fill this lack of a large ultradeep field on Coma, we
added several pointings from our previous deep multiband survey made
at CFHT (Adami et al. 2005 (A05), 2006 (a and b: A06a and b), and 2007
(A07)) which have a subarea in common. The resulting field of view of
this new dataset combination ($\sim 42\times$7~arcmin$^2$ or $\sim$300
arcmin$^2$) is large enough to study possible environmental effects
and deep enough (R$\sim$25.5) to reach the very faint object
populations in the globular cluster regime.

Section 2 presents our data and methods. Luminosity functions are
computed in Section 3. We discuss the nature of the faintest
objects statistically in the Coma cluster in Section 4 and 
give conclusions in Section 5.
All along that paper we use the following cosmological parameters: 
H$_0$ = 71 km~s$^{-1}$~Mpc$^{-1}$,  $\Omega _\Lambda =0.73$ and 
$\Omega _m=0.27$. The resulting Coma distance modulus is 34.98.

\section{Data}

\subsection{Images and catalogs}

The Coma cluster images are described in detail in A06a. The set of
data was made of two deep 42$\times$30~arcmin$^2$ fields observed at
the CFHT 3.6 m telescope with the CFH12K camera: one covering the
north and the other the south part of Coma. Each field was observed in
the B, V, R, and I CFHT filters (u* band data is presently being
acquired with the Megacam camera) and magnitudes were derived in the
Vega system.

These observations had an overlap of $\sim 42 \times 7$~arcmin$^2$
in the form of a horizontal central strip (CS) crossing the Coma
cluster from East to West. The CS covers the $\alpha$=[194.51,195.32]
and $\delta$=[27.93,28.06] rectangle. In the CS the exposure time was
therefore at least doubled compared to the A06a data. We took
advantage of this to compute deep B, V, and I images. We also computed
a very deep R band image in the CS by co-adding (in addition to the north
and south fields) a south R CFH12K image not used in A06a. The
resulting exposure times in the CS are given in Table~\ref{tab:area}. Only 
the western field of Fig.~\ref{fig:Fig3} in R has a lower exposure time of 
10800sec due to a CCD not available during the observations of the additionnal
south R CFH12K image.

\begin{table}
\caption{Exposure times (for the R band: whole CS / western field),
  catalog 
(conservative) $\sim$100$\%$
completeness magnitudes (for the R band: whole CS / western field), and seeing 
(in arcsec) in the central $\sim 42 \times$7~arcmin$^2$ strip.}
\begin{tabular}{cccc}
\hline
Band & Exposure time (s) & Completeness magnitudes & Seeing \\
\hline
B & 14400 & 25.25 & 1.07 \\
V &  7440 & 24.75 & 1.00 \\
R & 14100/10800 & 25/24.7 & 0.90 \\
I & 14400 & 24 & 0.90 \\
\hline
\end{tabular}
\label{tab:area}
\end{table}

\begin{figure} 
\centering
\mbox{\psfig{figure=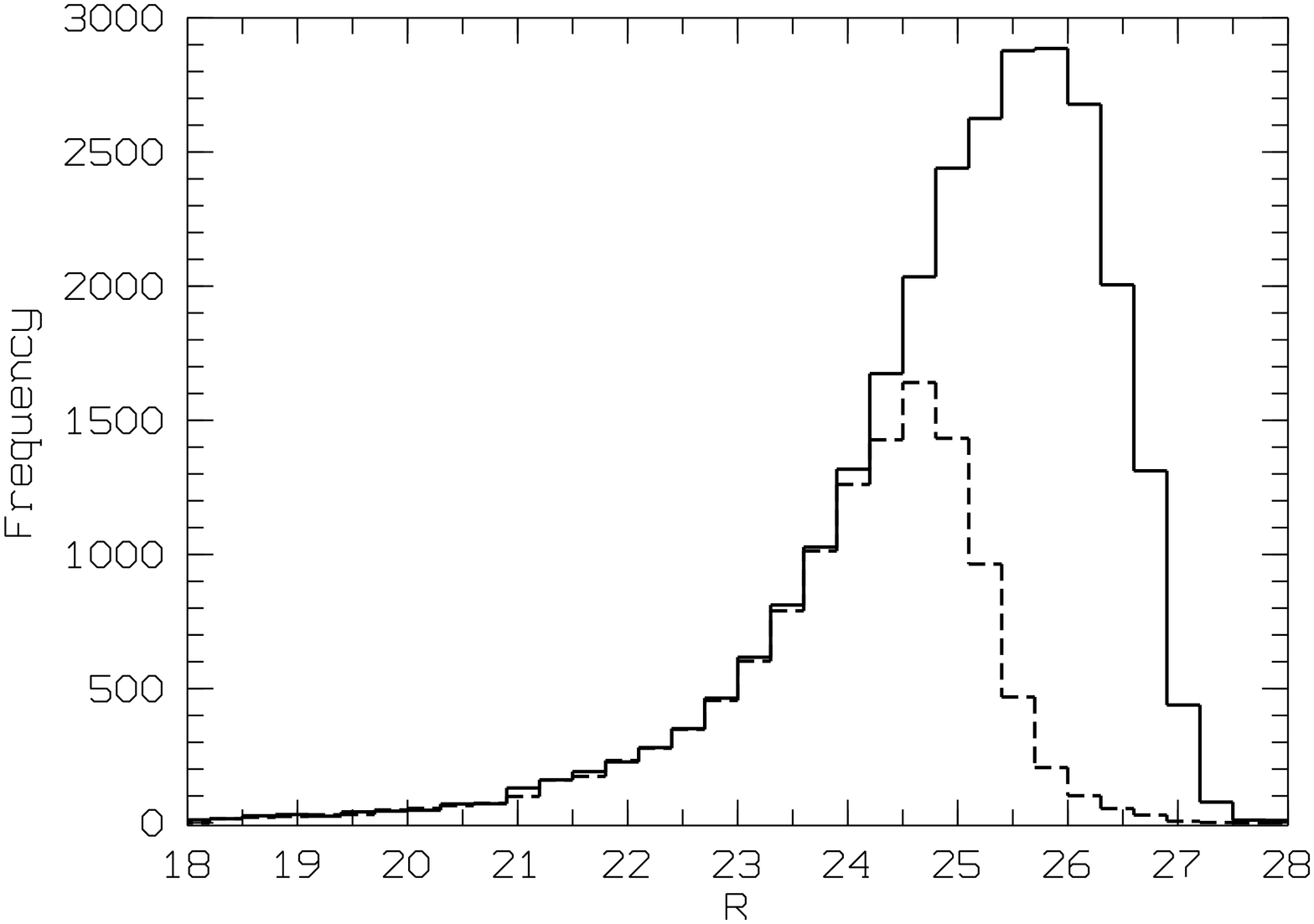,width=8cm,angle=270}}
\mbox{\psfig{figure=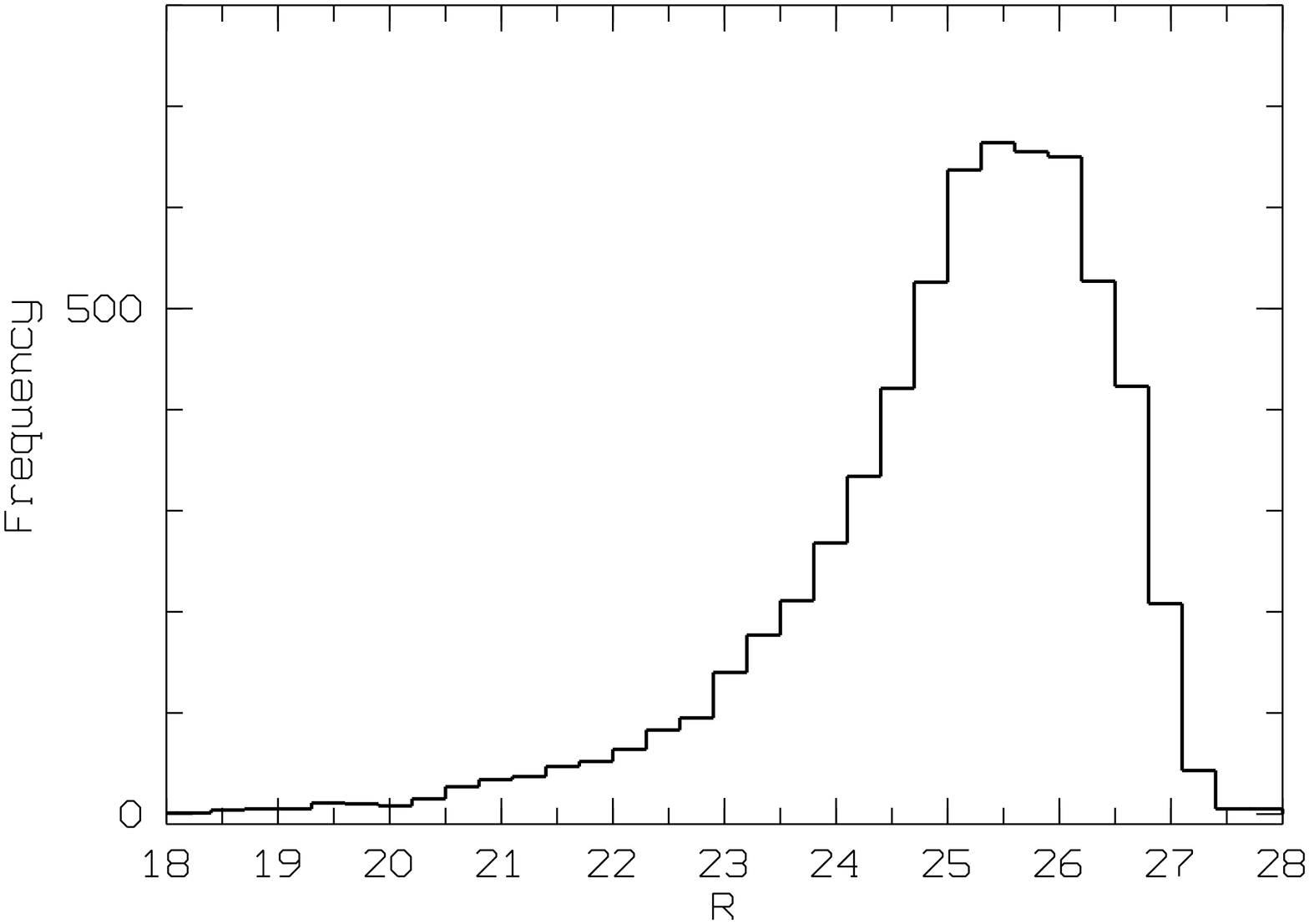,width=8cm,angle=270}}
\caption[]{Upper figure: Solid line: R band magnitude histogram for all the detected
  objects in the non masked regions of the CS. Dashed line: same in A07.
Lower figure: western field counts in the non masked regions
} 
\label{fig:Fig1} 
\end{figure}

We extracted catalogs of objects using the SExtractor package (Bertin
$\&$ Arnouts 1995) as in A06a. However, we chose a lower threshold and
minimal significant area: at least 3 pixels above a 1.1 sigma level,
in order to detect objects with the lowest possible S/N. We also
removed from the sample all objects with a size given by SExtractor
smaller than 1 pixel (0.206 arcsec).  We show in Fig.~\ref{fig:Fig1}
the resulting R-band magnitude histograms.  The histogram peaks
at R=25.5 for the whole CS R band image. However, the western field
has an exposure time lower by 30$\%$. We therefore also plot the
western field counts in Fig.~\ref{fig:Fig1}; this confirms that the
western field is slightly shallower. We also show in
Fig.~\ref{fig:Fig1} the A07 counts in the CS region. We will discuss
the completeness levels more precisely in section 2.3. 

We plot in Fig.~\ref{fig:Fig2} the R central surface brightness
($\mu_R$) versus the R band total magnitude for the present catalogs
(whole CS and western field of the CS) and the A06a catalog. In
addition to Fig.~\ref{fig:Fig1}, this illustrates well the gain in
depth for both total magnitudes and $\mu_R$ compared to A06a. To
be conservative, we chose to limit all the present catalogs to
$\mu_R$=25.7 to be sure not to undersubtract the field contribution.

\begin{figure} 
\centering
%\mbox{\psfig{figure=RmuR.ps,width=8cm,angle=270}}
%\mbox{\psfig{figure=RmuR2.ps,width=8cm,angle=270}}
\caption[]{R central surface brightness versus R total magnitude. The green lines are
used to estimate the completeness level of the samples (see text). 
Upper figure: whole CS in the non masked regions with red dots: A06a catalog
 and black dots: present catalog. The inclined black line at bright magnitudes shows
the separation between stars and galaxies. Lower figure: western field 
objects in the non masked regions.}
\label{fig:Fig2} 
\end{figure}

Finally, we applied  the same masking pattern as in A07 to
avoid CCD defects and regions heavily polluted by diffuse light from 
objects brighter than R=18 in the surrounding region (areas of 2 times 
the radius of these objects were removed from the data).

We compared our Coma data with a 1deg$^2$ field already described in
A07 and McCracken et al. (2003): the so-called F02 field. This
field was observed with the same telescope and camera, in the same
magnitude bands, and with similar exposure times and seeing, as
already stated in A07. We did not use the other comparison field
(F10) because it was not deep enough in all magnitude bands compared
to the present Coma cluster data. New catalogs were extracted
from the F02 field using the same extraction parameters previously
described, in order to be homogeneous with the present Coma catalogs.

\subsection{Star galaxy separation}

Star-galaxy separation was made like in A06a, from the central surface
brightness versus total magnitude diagram (Fig.~\ref{fig:Fig2}). We
classified as stars all objects brighter than R=21.2 and with a
central surface brightness brighter than 0.95R + 1.35 (these objects
are already removed from Fig.~\ref{fig:Fig2} where we only show the
separation line between galaxies and stars).

We show in A06a that this method allows to efficiently remove stars
from the catalogs down to R$\sim$21, as it optimally separates the
star and galaxy sequences in the total magnitude versus central
surface brightness plot. Fainter objects were assumed to be galaxies
(or globular clusters) because we also showed that the star
contribution becomes very small at R$\geq$21.

The star-galaxy separation has, however, a minor influence on our results
since the bright end of our luminosity function will be chosen at R=21 
(see below).

\subsection{Completeness}

A usual way to estimate the completeness level of a catalog is by
comparison to a significantly deeper catalog. In A06a, we compared for
example our data to the B95 catalog. Such a comparison is
not efficient here because the R band CS image is now at least as deep
as the B95 image and we would need a significantly deeper
reference. The M07 catalog offers such a deeper catalog but the field
of view in common with our data is far too small to give robust
results.

We apply instead the same method as in A07 (see their Fig.~6), based
on the total magnitude versus central surface brightness diagrams.  In
a few words, we first consider a magnitude range bright enough to be
sure to detect all objects (galaxies, stars, and globular clusters) 
whatever their surface brightness
(R$\leq$23.5).  We then consider the upper envelope of the central
surface brightness of the object population as a function of total
magnitude (inclined green lines in Fig.~\ref{fig:Fig2} for the R
band). Second, we define the central surface brightness limit of our
survey (horizontal green lines in Fig.~\ref{fig:Fig2}). As shown in
A07, the intersection between these two lines gives a good (and
conservative for compact objects as elliptical galaxies or globular 
clusters) estimate of the catalog completeness level whatever the
object type. This was done both for the whole CS and for the 
western field. These levels are summarized in
Table~\ref{tab:area}. We note that the R band image is complete down
to fainter magnitudes for compact objects (around R$\sim$25.5 as
suggested by Fig.~\ref{fig:Fig1}).

\section{Luminosity functions}

We limited our study to the very faint part of the LF (R$\geq$21). 
Brighter magnitudes were extensively investigated in A07 and
we want to focus here on objects that are potentially created in the cluster 
from material expelled from already existing galaxies. A candidate
population is tidal dwarf galaxies and the limit of R=21 corresponds 
approximately to their bright end (A06b).

We also only computed LFs in the R band because it is the only band
reaching a depth of R=25, and comparable to ultra-deep surveys such as M07
or B95. The other bands (B and I) will only help to characterise
the faint Coma cluster object populations.

\subsection{Luminosity function computation}

We have the same statistical approach as in A07. Briefly this consists
in estimating the fore and background contributions along the Coma
line of sight by using a comparison field. We then have access to the
number of galaxies inside the Coma cluster as a function of R
magnitude N$_{Coma\ cluster}$(R) with the following expression:

N$_{Coma\ cluster}$(R) = N$_{Coma\ line\ of\ sight}$(R) -
  N$_{empty\ field}$(R)

\noindent
where N$_{Coma\ line\ of\ sight}$(R) and N$_{empty\ field}$(R) are the
numbers of galaxies along the Coma cluster and along the empty field
lines of sight, respectively, as a function of R magnitude.

Error bars on the counts were computed with the same method as in A07,
taking into account magnitude uncertainties, Poisson noise and cosmic
variance. Cosmic variance was treated using the Huang et
al. (1997) formalism with the galaxy correlation function and the
field galaxy luminosity function in the F02 comparison field as entry
parameters.

\subsection{Results}

\begin{figure*} 
\centering
\mbox{\psfig{figure=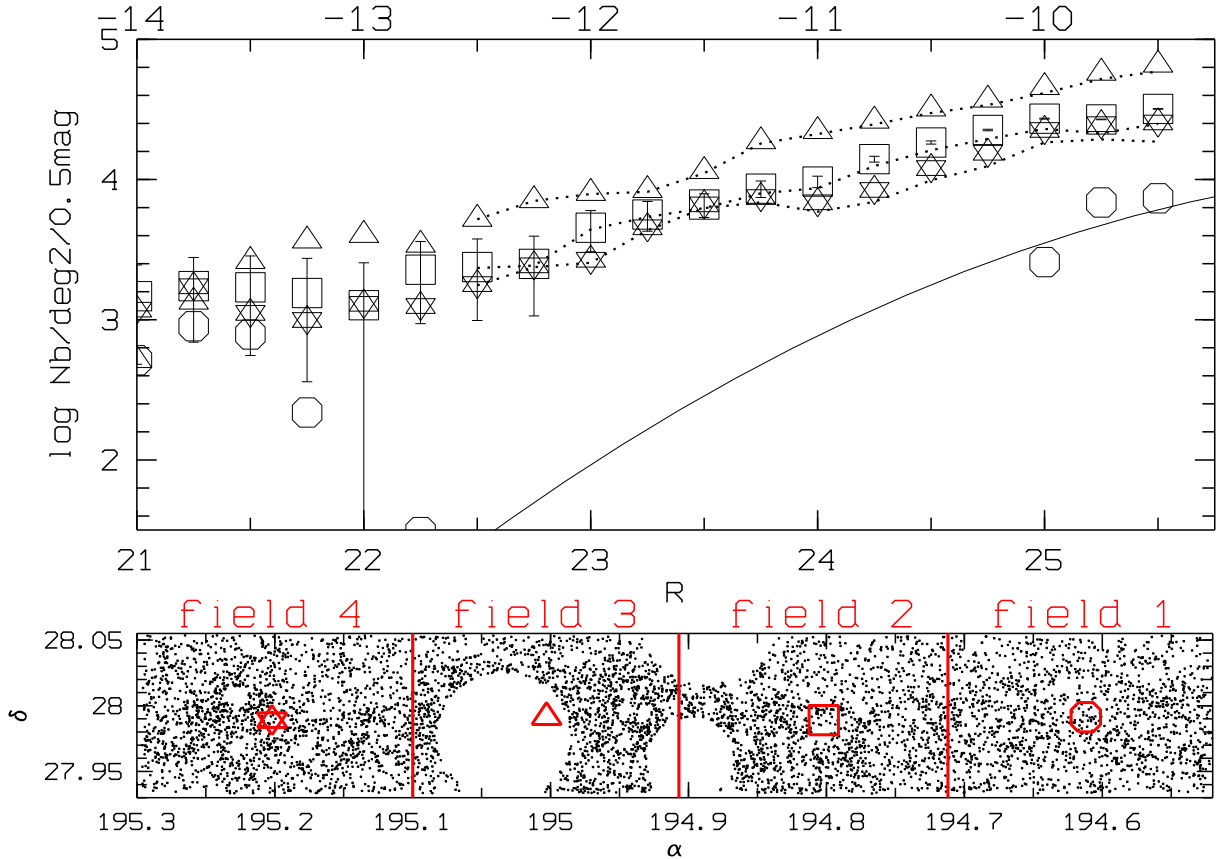,width=17cm,angle=270}}
\caption[]{Lower figure: R detected objects (stars being excluded)
down to R=25.5 outside of the masked regions. The four subfields are
delimited by vertical red lines. For clarity, only 40$\%$ of the
objects (randomly selected) have been plotted.  Upper figure: LFs in the four
corresponding subfields. Error bars are shown only for the East subfield
for clarity. Continuous line: predicted numbers of Globular Clusters
given the applied masking (see text). Dotted lines are the LFs
computed after subtracting the Globular Cluster counts from the galaxy
counts.  Symbols of the upper figure are shown in the corresponding
subfields in the lower figure. Upper magnitude axis is computed applying the
34.98 distance modulus.}
\label{fig:Fig3} 
\end{figure*}

We arbitrarily divided the CS into 4 sub-regions of 10~arcmin width in
right ascension to study environmental effects. This was a compromise
between the field size and population richness.  The four regions are
drawn in Fig.~\ref{fig:Fig3} as well as the objects detected down to
R=25.5. The Coma cluster LFs for the four subfields are displayed in
the same figure.

The general behavior of the LFs is similar to the results shown in
A07: a very poor LF in the western field and growing and richly
populated LFs in the other three fields. The fact that the
western field has a shallower completeness level 
%magnitude 
does not affect this conclusion, since the western field LF is still very
poor for R brighter than 24.7, where the data 
%in this field 
are complete. The western field only shows numerous objects at the
faint end of the LF.  The three other fields exhibit LFs with a very
regular slope. Before globular cluster subtraction, a power law model
gives slopes of -1.9$\pm$0.05, -2.0$\pm$0.05, and -1.9$\pm$0.05 from
East to West respectively; these slopes are quite similar to that
given by M07 for data of similar depth in a very central region of the
cluster ($-2.29 \pm$0.33), but steeper than the slope fitted by B95
($-1.42 \pm$0.05). The comparison field used in this paper is
significantly larger than those used by M07 and B95; therefore, we
expect our slopes to be more reliable than those obtained in previous
studies. The values of our slopes are different from what was seen for
magnitudes brighter than R=21 in A07 where several dips were
detected. We will discuss below this difference in terms of external
processes acting on galaxies. We note that the bright parts of the CS
LFs connect well with the faint parts of the A07 LFs in comparable
areas.

%{\bf They are not exactly the same
%because the four CS areas do not completely match the A07 regions and
%because the present catalog is significantly deeper in terms of total
%magnitude and surface brightness. Populations of low surface brightness
%objects that were not included in the A07 catalogs are therefore detected 
%in the CS.}

\section{Discussion}

\subsection{Nature of the faintest Coma cluster objects: are they globular
  clusters?}

We detected a very large population of objects as faint as M$_R=-9.5$.
Globular Clusters (hereafter GCs) have comparable magnitudes and we
must first estimate the intergalactic GC contribution to these counts
as the question of the nature of the faintest cluster objects is
recurrent. 
%We try here to check if the hypothesis of a faint Coma cluster object
%population dominated by GCs could be excluded.
We discuss here the hypothesis of a faint Coma cluster population 
dominated by GCs.

Our data are, by definition, free of GCs close to their parent bright
galaxies because we rejected all regions closer than twice the
radius of all galaxies brighter than R=18. This strongly limits the GC
contamination shown for example in B95. However, intergalactic GCs or
GCs distant from galaxy centres can still be present, as demonstrated
by Rhode $\&$ Zepf (2004: RZ04).  According to these authors, the GC
number density profile could have a small (but non negligible)
contribution to number counts outside the regions masked in our
analysis. 

If we apply our masking to the RZ04 M86 data, we find that the
remaining GCs (distant from their parent galaxy) are $\sim$80 times
less numerous than the whole population.  In order to estimate a GC LF
in our fields, we assumed the M86 population of GCs to be typical of
all galaxies brighter than R=18 in the Coma CS. It will only be used to
give us a rough estimate of the GC LF. Then,
rescaling the surface densities to the Coma cluster distance, we were
able to generate the GC LF shown by the continuous line in
Fig.~\ref{fig:Fig3}.  This estimate neglects the GC population coming
from galaxies fainter than R=18. However, galaxies just brighter than
R=18, and therefore very significantly fainter than M86, also have a
lower GC richness compared to this galaxy. This compensates at least
partially the first underestimate.

On the one hand, we show that GCs can explain a large part of the
object counts in the western field (Fig.~\ref{fig:Fig3}). This
does not mean that this field is only populated by intergalactic GCs
at magnitudes fainter than R=25, but that we can not exclude that a
large part of the faint object population in this field could be
GCs. We have, however, to take into account the fact that at these
faint magnitudes, the western field data are becoming incomplete and
that the easiest objects to detect are globular clusters because they
have a compact shape.

On the other hand, the three other fields have a faint object
population that is clearly more numerous than the population of GCs
predicted using the M86 data. This probbaly means that these three
fields do not have a high level of GCs compared to their overall very
faint galaxy population.  Subtracting the estimated GC LF to the
overall LFs does not even allow to completely flatten the slope (see
dotted lines in Fig.~\ref{fig:Fig3}). This is consistent with the
results of A07 who have shown that only the NGC~4874 field has a
significant population of objects that could be intergalactic
GCs. Evidence for this last significant population was only given for
B$\leq$24.75 and our B and V data are not deep enough to generalize
this study down to R=25.5.

In conclusion, the very faint object population in the three eastern
fields is probably dominated by galaxies.

\subsection{Faint end of the galaxy luminosity functions}

We confirm (compared to A07) that the western part of the Coma cluster
is very poorly populated, with less than 30 objects per deg$^2$ and
per half magnitude bin down to R=25. Even for fainter magnitudes, the
level remains lower (around 4000 objects per deg$^2$ and per half
magnitude) than the populations in the other fields. We have previously 
shown that these objects are
unlikely to be galaxies for a large part. This subfield is included in
what we called the South area in A07 and it was probably at least partially
depopulated in terms of galaxies by major matter infalls coming from
the West (see also A07).

LFs in the three other subfields are confirmed (compared to A07) to be
strongly rising with slopes close to $-2$. This implies the existence of
a very large population of dwarf galaxies in these regions and shows that
the LF in the central part of the Coma cluster is very significantly
different from the field LFs (e.g. Ilbert et al. 2005). This also
shows that peculiar processes are at work to form or to concentrate at
this place such a large amount of faint objects.

We know that at least part of the faint Coma cluster dwarves (low
surface brightness objects, see A06b) were probably formed in the
early stages of the cluster, so at least part of the faint galaxy
population is made of old cluster-resident galaxies. We have also
shown in A06b that another part of these low surface brightness
objects does not share the same history and are possibly coming from
the field. A possible scenario in agreement with the present data
could be that debris from brighter disrupted and harassed galaxies are
populating the very faint part of the Coma cluster galaxy LFs. The
mass spectrum of these debris would have a slope close to $-2$. This
could be verified via numerical simulations, testing the mass
distribution of debris issued from galaxy close encounters,
generalizing for example the results by Bournaud et al. (2003). An example of
such candidates is seen in our data around the two possibly 
interacting galaxies shown in Fig.~\ref{fig:exa}.

Other concentrations of stars, which are called knots, have
also been observed to be ejected from spiral galaxies in clusters by
Cortese et al. (2007). The faintest knots observed by these authors
from HST data have magnitudes F$_{475}=-11.5$ (close to our B band),
and are therefore somewhat brighter than the faintest of our objects,
but not very far in magnitude (and therefore mass) range. They also are very
blue, so are still forming stars. Similar blue knots are also observed in our 
data for example south-west of NGC4858 (see Fig.~\ref{fig:exa}) with 
magnitudes as faint as R$\sim$24.5, but it is besides the goals of 
this paper to discuss a precise scenario of what happens around this galaxy.  

\begin{figure*} 
\centering
%\mbox{\psfig{figure=example.ps,width=12cm,angle=270}}
%\mbox{\psfig{figure=ngc48604858.ps,width=12cm,angle=270}}
\caption[]{Trichromic B, V, and R band color images. 
  Upper figure: possibly interacting galaxies in the Coma cluster. Coordinates of the 
  two close galaxies surrounded by faint objects at
  the upper left of the figure are 13 00 06.18, 28 15 05.8. The field size is
  1.9'$\times$1.7'.
  Lower figure: NGC4858 and NGC 4860. A concentration of blue objects is
  observed south-west of NGC4858. The field size is
  3.8'$\times$3.2'.}
\label{fig:exa} 
\end{figure*}

The fact that the three eastern field LFs are very regular (except
perhaps around R=22) shows that the general processes proposed in A07
to explain the dips in the LFs detected down to V$\sim$22 are not
efficient on the Coma very faint galaxy population.  Similarly, the
two central fields (including the two Coma cluster dominant galaxies
NGC~4874 and NGC~4889) are the most populated regions. This shows that
galaxy disruption sometimes proposed to explain the lack of faint
galaxies in the cluster centers does not act too strongly on the
faintest existing galaxies (or knots). The large scale diffuse light
sources detected in A06b around NGC~4874 then probably originate
from brighter galaxies.

\subsection{Nature of the faintest Coma cluster galaxies}

Besides the fact that these galaxies could be (at least part of them) debris of
larger objects, we investigate here their properties based on colour
plots such as B-I versus B-R (see A07). 
Ideally, we would need B and I data of similar depth as the R data,
but this is not the case. We will therefore limit our sample to
B=25.25 (the B band conservative completeness level). This means that all 
objects detected in B will also be detected in R and I, assuming 
typical B-R and B-I colours. 

We then compute galaxy density maps in the B-I/B-R space for the Coma
cluster line of sight and for the F02 comparaison field, as already
done in A07. The difference between the two maps gives the statistical
distribution of objects inside the Coma cluster CS in a B-I versus B-R
plot.

In such a plot, we have shown (A07) that early-type, early-spiral
and late spiral galaxies were optimally separated in the B-I versus
B-R space by the lines:

(B-I) = $-1.2$ (B-R) + 1.45

\noindent
and

(B-I) = $-1.2$ (B-R) + 2.60

With these limits, only 15$\%$ of the Coleman et al. (1980) type 2
(Sbc) galaxies are misclassified as type 1 (E/S0), and only 26$\%$ of
types 3/4 (Scd/Irr) are classified as type 2. Results are displayed in
Fig.~\ref{fig:Fig4}.

\begin{figure} 
\centering
%\mbox{\psfig{figure=field1.ps,width=7cm,angle=270}}
%\mbox{\psfig{figure=field2.ps,width=7cm,angle=270}}
%\mbox{\psfig{figure=field3.ps,width=7cm,angle=270}}
%\mbox{\psfig{figure=field4.ps,width=7cm,angle=270}}
\caption[]{Coma cluster galaxy density in the B-I/B-R space. Density
is proportional to the blackness of the shaded regions. We only show
the regions significant at more than the 3$\sigma$ level. The two
black inclined lines delimit the separation between early-type (upper
right of the figures), early-spiral (between the two lines) and late
spiral galaxies (lower left of the figures). The horizontal lines are
the mean B-R colour for GCs as from RZ04 along with the approximate
error envelope (horizontal dashed-lines) of $\pm$0.1. From top to
bottom: West to East.}
\label{fig:Fig4} 
\end{figure}

We can note systematic concentrations around B-R$\sim$1.2 and
B-I$\sim$1.7 whatever the field. These concentrations correspond
nearly exactly to the expected B-R GC location from RZ04. Moreover,
the GC B-I colours are probably close to bright elliptical galaxy
colours, so are located above the (B-I) = $-1.$2 (B-R) + 2.60
line. This is also where the previous concentrations were found and
they are probably originating at least partially from GCs. For
the western field, however, these possible GCs are clearly dominated
in population by the other objects: they only represent 8$\%$ of 
the total number. This leads us to conclude that the 
%not-too-faint 
objects with B brighter than 25.25 are probably not all GCs. So, only
the faintest part of the western field object LF could be really
dominated by GCs, as also shown in Fig.~\ref{fig:Fig3}, assuming that
the western field incompleteness and the compact GC shape do not
conspire to increase artificially the GC contribution (as discussed in
section 4.1).  

The westernmost field (field 1) appears dominated by late type galaxies
that could be infalling objects. This is not surprising, since it is
located on the path of infalling structures onto the Coma cluster
(e.g. Neumann et al. 2003) which are likely to be populated by late
type field galaxies.

The next field (field 2) to the East is dominated by NGC~4874 and
exhibits late type faint galaxy populations. This is in good agreement
with A07 (see their Fig.~14), where early type galaxies become
dominated by later type objects above R$\sim$22.75.  The next field
(field 3) to the East is dominated by NGC~4874 and NGC~4889 and has an
earlier type population compared to field 2, but still made of
relatively late type objects. This was also visible in Fig.~14 of A07.
It is quite puzzling to find these galaxy types around the dominant
cluster galaxies except if these are the remnants of disrupted disks,
so still forming stars. These galaxies could be similar to the blue
low surface brightness galaxies already detected in A06b, that were
found to be clustered around NGC~4889 and present in the western
areas. Only very high resolution numerical simulations could
confirm or not this statement.

The easternmost field (field 4) has a mixed population of all types, but
dominated by early types, as expected in the absence of major
infall. Only the redder objects in B-I could be globular clusters
because most of these objects do not have sufficiently red B-R colours.

\section{Conclusions}

We first note that the present study has mainly used R band data to
detect the objects and B and I band data to investigate their
nature. The fact that the B band data are not as deep as those
in the R band is clearly the limitating factor of our work. Let
us now summarize our main findings:

- Field 1 is very poorly populated. This is possibly due to a
  combination of galaxy destruction and to the fact that infalling galaxies do
  not stay in this region, as suggested in A07.

- The LF slopes (close to $-2$) are very different in the CS compared to field 
regions free from rich clusters (e.g. Ilbert et al. 2005). 

- No significant dips are detected for the very faint population showing that
  peculiar processes creating such dips in brighter galaxy populations (see
  A07) are not efficient here. 

- The western faint galaxies are mainly blue, and therefore
  late-type-like or issued from late-type galaxies included in the
  West infalling galaxy layers (see Neumann et al. 2003).

- The eastern galaxies do not show preferential colours so do not have a
  dominant component. This is probably related to the absence of major
  infalls in this region.   

- The two areas dominated by NGC~4874 and NGC~4889 are preferentially
  blue, and therefore late-type objects, and possibly come from disk-like
  disrupted disks.

The main insight of this work compared to A07 on the cluster formation
history is the absence of significant dips in the very faint galaxy
LF.  A possible scenario in agreement with the present data could be
that debris from disrupted and harassed brighter galaxies are
populating the very faint part of the Coma cluster galaxy LFs. The
mass spectrum of these debris would have a slope close to $-2$ but
this still needs to be verified with numerical simulations.

This property could also be directly related to the fact that the CS
LFs are very different from the field LFs.  Milne et al. (2007)
reached similar conclusions. To explain this difference, one needs to
flatten the field LF faint end or to steepen the cluster LF.  A fading
of the faint field galaxies due to reionization processes alone
(Benson et al. 2003) cannot explain this difference, as already noted
by Milne et al. (2007). They proposed, therefore, other environment
dependent feedback processes. However, the cluster LF can be steepened
when the faint magnitude ranges are populated by debris coming from
brighter galaxies (such as those observed by Cortese et al. 2007 for
example), as proposed in the present work.

\begin{acknowledgements}
The authors thank the referee for constructive comments. \\
The authors are grateful to the CFHT and Terapix teams for their help
and to the French PNG, CNRS for financial support.
\end{acknowledgements}

\end{document}